# Mixed Organic Cation in Chiral Two-Dimensional Organic-Inorganic Hybrid Metal Halides – An ab-initio Study of Nonlinear Optical (NLO) Properties

Xiyue Cheng[1‡], S. Muthukrishnan[2,3‡], Hanxiang Mi[1], Shuiquan Deng[1*], Goncagul Serdaroglu[4], R. Vidya[2,3], and Alessandro Stroppa[5*]

[1] State Key Laboratory of Structural Chemistry, Fujian Institute of Research on the Structure of Matter (FJIRSM), Chinese Academy of Sciences (CAS), Fuzhou, 350002, P. R. China.

[2] Computational Laboratory for Multifunctional Materials (CoLaMM), Department of Physics, Anna University, Sardar Patel Road, Guindy, Chennai 600025, India.

[3] Centre for Materials Informatics (C-MaIn), Sir C V Raman Science Block, Anna University, Sardar Patel Road, Guindy, Chennai 600025, India.

[4] Sivas Cumhuriyet University, Faculty of Education, Math. and Sci. Edu., Sivas, Turkey.

[5] CNR-SPIN, c/o Dip.to di Scienze Fisiche e Chimiche - Via Vetoio - 67100 - Coppito (AQ), Italy.

*KEYWORDS: hybrid organic-inorganic perovskites, density functional theory, second harmonic generation, electronic structures, atom response theory*

**ABSTRACT:** The mixing of organic cations represents yet another direction to explore in the field of chiral *organic-inorganic hybrid metal halides* (OIHMH). Here, we perform structural optimizations, electronic structures, and non-linear optical (NLO) studies using the density functional theory of two recently synthesized chiral OIHMHs, [R-MePEA][C3A]PbBr$_4$, and [R-MePEA][C4A]PbBr$_4$, with mixed chiral arylammonium and achiral alkylammonium cations. We find that the noncovalent weak interactions (e.g. Br⋯NH interactions) play an important role in the formation of these OIHMHs. Our study further indicates that the two non-centrosymmetric compounds exhibit relative wide bandgaps (~3.5 eV), strong second harmonic generation (SHG) responses (~0.5-1.5×KDP), and moderate birefringence (~0.088), indicating possible applications NLO materials. Atom response theory analysis reveals that the SHG responses are determined mainly by the occupied Br *3p* non-bonding orbitals as well as by the unoccupied Pb *5p* orbitals which shows the important contribution of the inorganic PbBr$_4$ layer to the nonlinear optical properties.

1. Introduction

Chirality is a unique geometrical phenomenon that received great importance in the fields of physics, chemistry and materials science. If a specific molecule cannot overlap with its mirror image by proper symmetry operations, such as translations and/or rotations, the molecule is chiral, and the mirror images are called enantiomers. The research on chiral systems is mainly related to organic synthesis and applications in medicinal chemistry[1,2] since critical molecules in the living body are in the chiral framework.[3] In the past, the chiral inversion pathway of poly (BBPFA) structure was investigated by the atomistic approach using the chirality-based variables to get the free energy profile in this conversion.[4] Furthermore, the parity-violation energy due to weak forces, $\Delta E_{PV}$, (total electronic energy difference of two enantiomers) and ECM (electronic chirality measure) relationship was reported and the authors suggested that the weak forces producing parity-violating effects can be critical to understanding the biological homochirality phenomenon.[5] The presence of chirality can provide considerable flexibility in the functional material design due to its non-centrosymmetric nature.

*Organic-inorganic hybrid metal halides* (OIHMHs) have great technological potential in the field of photovoltaics, LEDs, and other optoelectronic applications. Two-dimensional (2D) OIHMHs attracted much research interest, due to their unique characteristics, such as good thermal and chemical stabilities.[6] They have interesting electronic properties such as tunable bandgaps, long charge carrier lifetimes, and mobility, higher absorption coefficient and high photoluminescence.[7] By carefully choosing the proper organic spacer cation in 2D OIHMHs, the tuning of its electrical and optical properties is achieved.[8] Mixing more than one cation leads to a new possible strategy to engineer the optoelectronic properties of 2D OIHMHs.[9,10] In particular mixing more than one organic cation was recently explored by Yan *et al*.[11] who studied aromatic ammonium-based 2D OIHMHs, Qiu *et al* studied the Ruddlesdan-Popper phases on tin-based mixed organic spacer cation (n-butylamine and phenethylamine) for solar cell application.[12] Similarly, Lian *et al* fabricated the butylammonium and methylammonium mixed spacer cation 2D OIHMHs, oriented with a grain size larger than 1 μm.[13] Mixing different organic cations in 2D OIHMHs may significantly modify their properties, leading to a reduction in the trap state density, incrementing the solar cell efficiency, and structural stability.[14]

The mixing of the chiral organic cations with the achiral organic ones in order to "chiralize" a 2D OIHMHs is considered another emerging area of research.[15] The resulting chirality transfer between chiral organic cations to the framework results in chiral 2D OIHMHs thus allowing for chiral-opto-electronic applications and spintronic properties.[16] Kim *et al* studied the chiral-cation-induced spin selectivity at room temperature LEDs, where the organic layer injects the spin-polarized holes, into the halide perovskite.[17] Furthermore, the presence of chiral ligands in the lead halide framework, enables strong circularly polarized light emission and detection.[18] Ishii *et al* used helical lead halide 1D perovskite with chiral organic-cation-based photodiodes, for the detection of circularly polarized light. Here circularly polarized light detection with 1D perovskite has the highest polarization discrimination ratio of 25.4, which is larger than other conventional devices.[19] Yang *et al* synthesized the first 2D perovskite ferroelectrics, by incorporating homochiral cations into the 2D OIHMHs, and studied the material by crystal structure analysis, and circular dichroism spectra.[20] The good structural and chemical properties of the halide perovskites, enabled the strong application in the field of non-linear optics (NLO).[21,22] NLO materials play a crucial role in modern laser technology as electro-optical materials for optical modulators, Q-switches, detectors, and frequency converters because of their nonlinear interactions with incident light.

Recently Yan *et al* [23] experimentally synthesized the single layer thick 2D lead bromide HOIPs comprising of chiral aryl organic cation part (MePEA: methylphenethylammonuim,) and achiral alkyl cations (C3A: propylammonium or C4A: butylammonium). They studied the mixed alkyl-aryl cations with a 1:1 molar ratio and characterized the structure for the first time by analyzing the role of CH···π interaction in the structural formation of chiral mixed 2D HOIPs.

In this work, we study the structural, electronic structures and NLO properties of the chiral mixed cation 2D HOIPs, [R-MePEA][C3A]PbBr$_4$ and [R-MePEA][C4A]PbBr$_4$. [23] We reveal the important role of Br···NH noncovalent interactions between the inorganic and organic layer in the formation of the 2D layered HOIPs. Our study further suggests that both OIHMHs show large SHG responses (i.e., ~1.49 and 0.53 ×$d_{eff}^{KDP}$) among the reported lead-containing OIHMHs which implies their possible application as NLO materials. Besides, the origin of the large SHG response is found to be the inorganic groups rather than the organic parts. For completeness, in supplementary material (SM), we also study the optical rotatory dispersion (ORD), and electronic circular dichroism (ECD) spectra of the neutral and cationic MePEA enantiomers.

2. Computational Details

The structural and electronic properties of [R-MePEA][C3A]PbBr$_4$ and [R-MePEA][C4A]PbBr$_4$ were calculated by using the density functional theory (DFT) and the projector augmented wave (PAW) method implemented in the Vienna Ab initio Simulation Package (VASP).[24-26] The generalized gradient approximation (GGA) of the exchange-correlation potential in the form of the Perdew-Burke-Ernzerhof (PBE) was used throughout this work[27]. The employed PAW-PBE pseudopotentials of elements Pb, Br, C, N and H treat $5d^{10}6s^26p^2$, $4s^24p^5$, $2s^22p^2$, $2s^22p^3$ and $1s$ as the valence states, respectively[28]. The plane wave cutoff energy for the expansion of wave functions was set at 600 eV with dense $k$-point mesh, 7× 7 × 4 for both compounds. The conjugate-gradient algorithm as implemented in the VASP code was used in all structural relaxation. In this work, the atomic positions were all allowed to relax to minimize the internal forces while keeping the unit cell lattice constants to experimental values. Convergence criteria for the energy differences (0.1 meV) and stress tensors ($\leqslant$ 0.005 eV/Å) were achieved.

The second-order NLO properties were calculated by employing the "sum over states (SOS)" methods using the code (ARTAROP) that we have recently developed[29] based on the calculated electronic structures obtained from the VASP[24-26] optical modules. The SOS formalism for second-order susceptibility was derived by Aversa and Sipe[30] and later modified by Rashkeev et al[31,32] and Sharma et al.[33,34] As the DFT electronic structure calculations underestimate bandgaps, the scissor operation[35] was used to increase the energies of the unoccupied states from those of the occupied ones to have the HSE06[36-38] bandgaps. The powder SHG response, the effective $d_{eff}^p$, an average SHG coefficient over all possible orientations of the powder crystals, are estimated from the formula derived by Kurtz-Perry[39] and Cyvin et al[40] based on the calculated non-zero SHG tensor elements. To evaluate the individual atom and orbital contributions to the SHG components, the atom response theory (ART) based on partial response functional (PRF) was used (see SM).[27] Finally, the ORD and ECD spectroscopic computations of both neutral and cationic MePEA enantiomers were performed by G16W software,[41] at B3LYP/6-311++G(d,p).[42,43] Chem3D 17.1[44] program was used to calculate the connolly solvent excluded volumes of MePEA enantiomers and compared with the corresponding data to confirm the electronic structures. Gauss View 6.0.16 package[45] was used for analysis and illustration of the computed results (See SM).

3. Results and discussion

3.1. Structural optimization

The relaxed structure for [R-MePEA][C3A]PbBr$_4$, and [R-MePEA][C4A]PbBr$_4$ are shown in Figure 1. We also compared the experimental crystal structures with the calculated ones using Bilbao crystallographic server database[46] as presented in Table S1 showing the degree of lattice distortion, the maximum displacement between the atomic positions, the arithmetic means value, and the overall measurement of similarity. The calculated values of structure comparison (with the experimental values) for both compounds in the PBE-GGA and DFT-D2 correction suggest a good agreement of our optimized structure models with the experimental structures. After the relaxation, the bond lengths for Pb-Br, C-C, N-C, and N-H slightly increase compared to the experimental values at the PBE-GGA level, as shown in Table S2. Such an increase is further weakly enhanced when considering the van der Waals dispersion correction. Generally, the inclusion of weak non-covalent interactions leads to a better structural description in the organic and inorganic hybrid systems. Both compounds crystallize in the NCS space group $P2_1$ (Figure 1a,b). Within each unit cell, Pb atoms have one crystallographically unique site, whereas the Br, N, C, and H atoms occupy 4, 2, 12 and 24 independent crystallographic positions, respectively. All atoms are at special Wyckoff positions of 2*a*. The Pb atoms are coordinated with six Br atoms, of which two Br1 and two Br2 are located in the *ab* plane, Br3 and Br4 are along the *c* direction, forming distorted PbBr$_6$ octahedra (Figure 1c,e). Within the *ab* plane, neighbouring Pb-centered octahedra are connected by sharing the Br1 and Br2 atoms with bond lengths around 3.06 Å, leading to an inorganic PbBr$_4$ layer. For the ones along the *c* axis, the Pb-Br3 distances (~3.07 Å) are similar to that of the inner *ab* plane Pb-Br bonds, while they are all obviously shorter than Pb-Br4 distances (~3.27 Å) indicating weaker Pb-Br4 bonding interactions. The formation of such distorted PbBr$_6$ octahedra may induced by the second-order Jahn-Teller (SOJT) distortion.

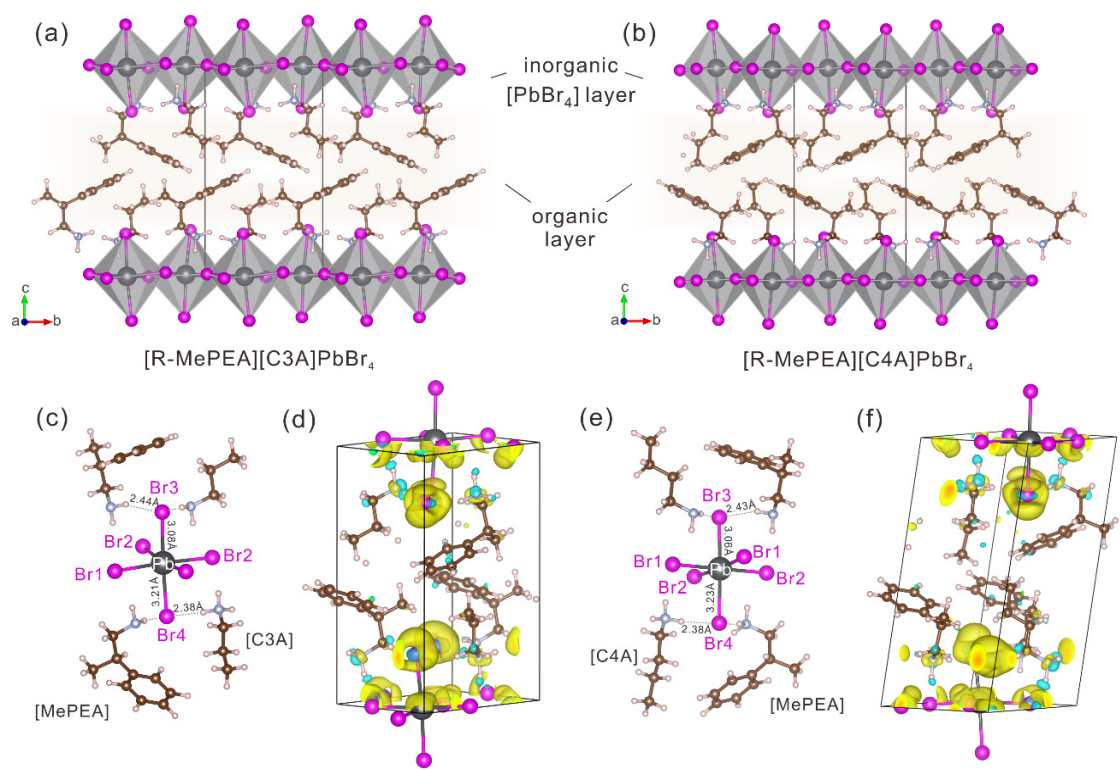

**Figure 1** 2D layered structures of [R-MePEA][C3A]PbBr$_4$ (a), and [R-MePEA][C4A]PbBr$_4$ (b). The coordination environment of PbBr$_6$ octahedron and the Br···NH weak interactions (c, d) together with the corresponding calculated charge density differences (d, f), for these two compounds, respectively.

The inorganic and organic layers are stabilized through noncovalent interactions, such as Br···NH interactions, which exist between Br atoms and the terminal NH$_3$ of R-MePEA cation and C3 or C4 alkyl cation, respectively.

The distance of Br⋯NH interaction in [R-MePEA][C3A]PbBr$_4$ and [R-MePEA][C4A]PbBr$_4$ are very similar, around 2.39-2.86 Å. These values are less than the maximum distance (3.20 Å) to have such weak interactions (1.05 times of the sum of the van der Waals radii of H and Br atoms) indicating the presence of Br⋯NH interactions. The noncovalent weak interactions are further evidenced by the calculated charge density differences (CDD) by using Δρ = ρ$_{compound}$ - ρ$_{inorganic}$ - ρ$_{organic}$, where ρ$_{compound}$ denotes the total charge density of [R-MePEA][C3A]PbBr$_4$ or [R-MePEA][C4A]PbBr$_4$, and ρ$_{inorganic}$ and ρ$_{organic}$ represents the charge density of inorganic [PbBr$_4$] layer and the organic mixed alkyl-aryl layer, respectively, as displayed in Figure 1d,f. The CDD results clearly show the inter-layer charge transition in both compounds in which the majority of charges transfer from the N-H groups of [R-MePEA] and [C3A] ([C4A]) to the PbBr$_6$ octahedra. Besides, the organic cations are also stabilized through weak interactions, such as CH⋯π interactions which were discussed in detail by Yan et.al [23]. It can be seen that the noncovalent weak interactions play an essential role in the formation of these 2D hybrid perovskites.

3. 2. Electronic structures

The electronic structures are described at the level of PBE-GGA including the spin-orbit coupling (SOC). The direct PBE bandgaps ($E_g^{PBE}$) at B=(0, 0, 0.5) for [R-MePEA][C3A]PbBr$_4$ and [R-MePEA][C4A]PbBr$_4$ are calculated as 2.75 and 2.76 eV, respectively. The SOC effect causes a band split of Pb-6p states at Z and D points which further reduces the $E_g^{PBE}$ to 1.98 and 1.97 eV ($E_g^{SOC}$), respectively (Figure 2a,b and Figure S1). Note that, generally both the PBE or PBE-SOC would yield band gaps relatively smaller with respect to the experimental gaps due to the well-known incorrect estimation of the quasi-particle energies, we also apply the hybrid functional HSE06 calculations with mixing parameter α=0.25 to obtain the $E_g^{HSE}$, 3.53 eV and 3.54 eV, respectively. For both compounds, the valence band maximum (VBM) is mainly dominated by the Br 4p and Pb 6s states, while the conduction band minimum (CBM) is primarily composed of unoccupied Pb 6p states. The C 2p, N 2p, and H p-like states from the organic part mainly occupied the VBs at ~2.5 eV below the Fermi level, while some relatively localized C 2p states lie at ~ 4.0 eV in the CBs. This clearly suggests that the bands near the Fermi level are dominated by orbitals from the inorganic PbBr$_4$ layer rather than the organic molecule in these two compounds.

The calculated partial density of states (PDOS) (Figure 2c, Figure S2a, and Figure S3) for both compounds show very similar features. The top valence bands (-2.3 eV to the Fermi level) consist of mainly Br 4p states and a weak mixing with C 2p states, while those between -15 to -2.3 eV are mainly composed of a large amount of hybridized C 2p and H s as well as N 2p and H s states. Relative weak Pb 5p valence states can be found around -2.0 eV, while the inner Pb 6s and 5d orbitals mainly localize at -7.0 and -16.0 eV, respectively. The bottom of the conduction bands ($E_g$-5.5 eV) is primarily made up of Pb-5p states and C-2p states, while the bands between 5.5 to 20 eV result from the unoccupied states of C, N, H, and Br atoms.

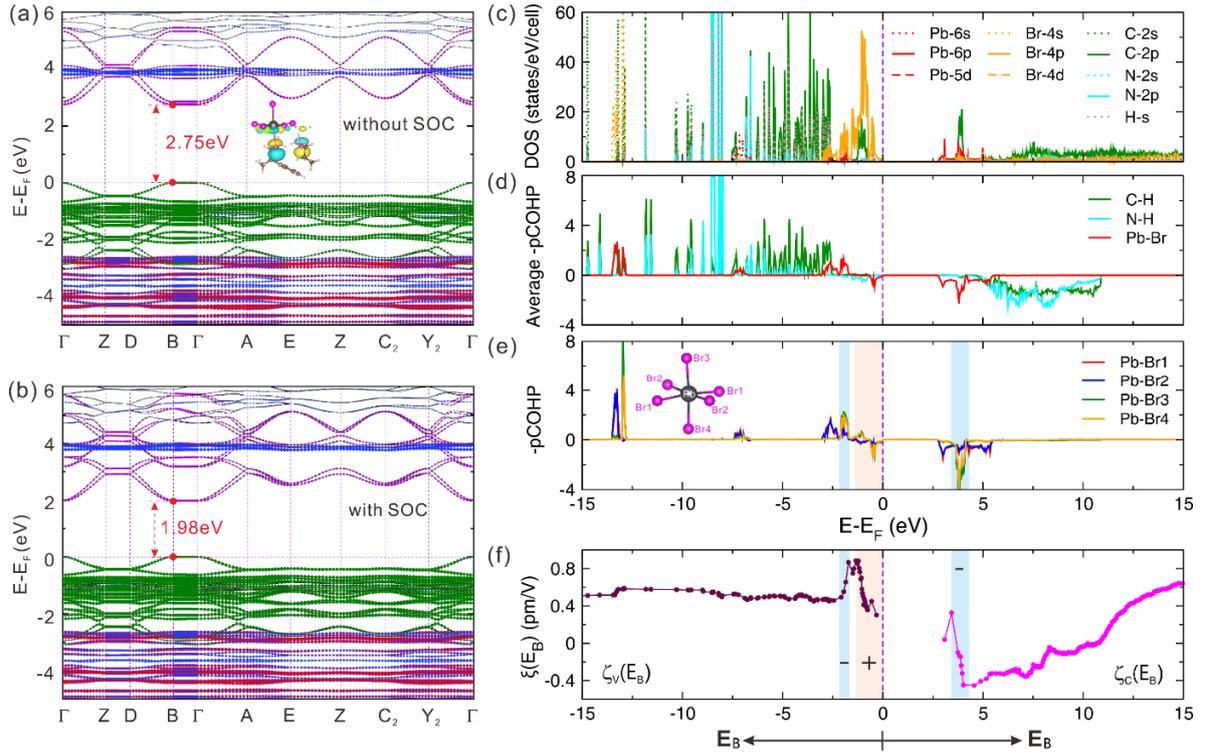

**Figure 2.** Band structures of [R-MePEA][C3A]PbBr$_4$ (a, b) without and with SOC. Γ=(0, 0, 0), Z=(0, 0.5, 0), D=(0, 0.5, 0.5), B=(0, 0, 0.5), A=(-0.5, 0, 0.5), E=(-0.5, 0.5, 0.5), C$_2$=(-0.5, 0.5, 0), Y$_2$=(-0.5, 0, 0). Key: green=Br 4p, purple=Pb 6p, blue=C 2p, red=H s. Insert picture is the wavefunction of VBM at B point. Calculated PDOS (c), COHP analysis describing the average chemical bonding (d) and PbBr$_6$ octahedron (e), as well as ζ($E_B$)-vs-$E_B$ plot for the largest SHG component d$_{25}$ (f) of [R-MePEA][C3A]PbBr$_4$.

## 3. 3. Bonding Interactions

To elucidate the bonding characteristics of the compound, crystal orbital Hamiltonian population (COHP)[47,48] analysis, had been performed (Figure 2d and Figure S2b). As expected, also the bonding characteristics, for both [R-MePEA][C3A]PbBr$_4$, and [R-MePEA][C4A]PbBr$_4$, are almost similar. The Br-4p states at the top valence bands are nonbonding states. Strong bonding interactions between the states of Br 4p and Pb 6p occurred in the region around -2.0 eV, while relatively weak interaction between Br 4p and Pb 6s can be found around ~ -7.0 eV. C-H and N-H p-s bonding interactions overlap over a wide energy range and the bonding strengths are found to be stronger than that of the Pb-Br interactions.

## 3. 4. Second Harmonic Generation (SHG)

Because of the point group of $C_2$, the SHG tensor for [R-MePEA][C3A]PbBr$_4$, and [R-MePEA][C4A]PbBr$_4$ has eight non-zero components, e.g. $d_{14}$, $d_{15}$, $d_{21}$, $d_{22}$, $d_{23}$, $d_{25}$, $d_{34}$ and d$_{36}$, as presented in Table 1. The Kleinman symmetry (KS) i.e., $d_{14} = d_{25} = d_{36}$, $d_{16} = d_{21}$ and $d_{23} = d_{34}$, is not strictly followed in these two compounds. The breaking down of the Kleinman symmetry means that the polarization directions of the mixed pump light and that of the exit light cannot be permuted, indicating the occurrence of dynamic anisotropy in the SHG response of the lower-symmetry structure[49] For this reason, KS is not enforced in calculating the NLO properties in this work. The static effective SHG value of [R-MePEA][C3A]PbBr$_4$, and [R-MePEA][C4A]PbBr$_4$ were calculated as 0.493, and 0.173 pm/V respectively (i.e., ~1.49 and 0.53 × $d_{\text{eff}}^{\text{KDP}}$). Note that the SHG response of [R-MePEA][C3A]PbBr$_4$ is stronger than that of Potassium Dihydrogen Phosphate (KDP, static $d_{\text{eff}}^{\text{KDP}}$ = 0.33 pm/V) and is comparable or stronger than that of some other OIHMHs , such as TMIM·PbCl$_3$ (powder SHG intensity $I_{\text{powder}}^{2\omega}$ ~1×KDP),[50] MHy$_2$PbBr$_4$ ( $I_{\text{powder}}^{2\omega}$ ~0.1×KDP),[51] MHyPbBr$_3$ ( $I_{\text{powder}}^{2\omega}$ ~0.18×KDP),[52] [CH$_3$(CH$_2$)$_3$NH$_3$]$_2$(CH$_3$NH$_3$)Pb$_2$Br$_7$ ( $I_{\text{powder}}^{2\omega}$ ~0.4×KDP)[53] and (MePEA)$_{1.5}$PbBr$_{3.5}$(DMSO)$_{0.5}$ ($d_{\text{eff}}$ = 0.68 pm/V).[54] Besides, the birefringence is also evaluated as the phase matching condition is important for the efficiency of an SHG process. The calculated birefringence value (Δn) for [R-MePEA][C3A]PbBr$_4$, and [R-MePEA][C4A]PbBr$_4$ are 0.089, 0.086 respectively at 1064 nm which fall in the range of technical applications, i.e. (0.03-0.10). Therefore, based on the calculated relatively wide band gaps, strong SHG responses, and moderate birefringence, it can be inferred that these two OIHMHs are potentially suitable for nonlinear optical applications, thus suggesting that further studies are needed in order to find mixed chiral OIHMHs with optimized NLO responses. The calculated specific rotation degrees [55-58] from 150-1064 nm and related ORD and ECD spectra for MePEA enantiomers were given in Table S3-S4 and Figure S4.

**Table 1. Calculated values of $E_g^{\text{PBE}}$, $E_g^{\text{SOC}}$, $E_g^{\text{HSE}}$, static SHG tensor $d_{ij}$, static effective response $d_{\text{eff}}$, and birefringence at 1064 nm.**

| Material | $E_g^{\text{PBE}}$ (eV) | $E_g^{\text{SOC}}$ (eV) | $E_g^{\text{HSE}}$ (eV) | $d_{ij}$ (pm/V) | $d_{\text{eff}}$ (pm/V) | Δn at 1064 nm |
|---|---|---|---|---|---|---|
| [R-MePEA][C3A]PbBr$_4$ | 2.75 | 1.98 | 3.53 | $d_{14}$= 0.50<br>$d_{16}$= -0.35<br>$d_{21}$= -0.27<br>$d_{22}$= 0.25<br>$d_{23}$= 0.26<br>$d_{25}$= 0.55<br>$d_{34}$= 0.26<br>$d_{36}$= 0.42 | 0.493<br>(1.49×$d_{\text{eff}}^{\text{KDP}}$) | 0.089 |
| [R-MePEA][C4A]PbBr$_4$ | 2.76 | 1.97 | 3.54 | $d_{14}$= 0.20<br>$d_{16}$= 0.08<br>$d_{21}$= 0.14<br>$d_{22}$= -0.09<br>$d_{23}$= -0.06<br>$d_{25}$= 0.21<br>$d_{34}$= -0.11<br>$d_{36}$= 0.14 | 0.173<br>(0.53×$d_{\text{eff}}^{\text{KDP}}$) | 0.086 |

The origin of the SHG responses is further investigated by employing the ART analysis.[27] The ART analysis enables one to determine the contributions of individual atomic orbitals, and hence those of individual atoms, to the SHG coefficients using the PRF method (see SM for details). Shown in Figure 2f and Figure S2c are the PRFs, $\zeta_V(E_B)$ and $\zeta_C(E_B)$. The $\zeta_V(E_B)$ functional increases in magnitude with decreasing $E_B$ from Fermi level to -1.5 eV indicating that the nonbonding Br 4p states contribute strongly to the SHG response in the VB part. Besides, the drastic change from CBM to 3.6 eV of the $\zeta_C(E_B)$ functional reveal the essential contribution from the unoccupied Pb-5p states. These results suggest that the SHG responses of these two HOIPs are determined largely by the occupied nonbonding orbitals Br 4p as well as by the unoccupied orbitals Pb 5p from the inorganic layer instead of the organic components.

The quantitative contribution $A_T$ (in %) of an individual atom τ to the strongest SHG coefficient $d_{25}$ for [R-MePEA][C3A]PbBr$_4$, and [R-MePEA][C4A]PbBr$_4$ are obtained based on the partial response functional (PRFs). As presented in Table 2, the $A_T$ of Pb (~19.9 %) in [R-MePEA][C3A]PbBr$_4$ is nearly 4.6 times that of Br (~4.3 %) and 14.1 times that of N (~1.4%). Besides, the averaged $A_T$ values of C and H are negligibly small, <0.4%. These results reflect that the Pb and Br serve as the NLO-active centers at the atomic scale. Considering the number of atoms of each element in the unit cell, the total contribution of Pb, Br, H, C, and N in [R-MePEA][C3A]PbBr$_4$ are 39.8, 34.5, 11.4, 8.7, and 5.6 %. Although the uneven stoichiometry effect is included, the metal atom Pb is still the leading contributor to SHG response, followed by Br atoms. An identical order of elemental contributions to the total SHG response, Pb > Br > H > C > N, can be seen in [R-MePEA][C4A]PbBr$_4$. For both compounds, the contribution of Pb atoms in the valence band can be attributed to the localized Pb 6s and 5d states, while the conduction band contribution mainly originates from the unoccupied Pb 5p states. These findings further quantitatively show that the SHG of [R-MePEA][C3A]PbBr$_4$, and [R-MePEA][C4A]PbBr$_4$ mainly originated from the occupied states of Br 4p, Pb 5s, and Pb 5d, and by the unoccupied states of Pb 5p.

Additionally, we note that the $A_T$ of Br atoms varies by their atomic sites. For both compounds (Table 2), the total contributions of SHG for Br1 and Br2 within *ab* plane are larger than those of Br3 and Br4 atoms along the *c* axis. This fact can be interpreted by analysis of their bonding status, PRF and PDOS, as shown in Figure 2d,f, Figure S2 and S3, respectively. In the top VBs (-1.5 to $E_F$), the nonbonding states from all Br atoms make a strong positive contribution to the SHG response. However, from -2.2 to -1.5 eV in VB where the Br3(Br4) 4p states make strong bonding interaction with Pb atoms, the PRF $\zeta_V(E_B)$ decrease sharply with the decreasing $E_B$. Similarly, a quick drop of $\zeta_C(E_B)$
functional can be found from 3.6 to 4.2 eV in CB where strong Pb-Br3(Br4) antibonding states dominate. Both features show opposite responses in these two energy ranges with respect to the PRF at top VBs or bottom CBs, indicating negative contributions from Br3 and Br4 atoms to the SHG response. Besides, in the bonding region related to Br1 and Br2 atoms, the $\zeta_V(E_B)$ functional change little which implies small contributions from these bonding states. Therefore, the total contribution of SHG for Br3 and Br4 is relatively smaller than that of Br1 and Br2. The reason for such negative contributions is not clear yet but it may relate to the orbital orientations and the Br⋯NH noncovalent interactions. Additionally, we found the individual SHG contribution of Br3 atom is lower than that of Br4 in both compounds. This may caused by the fact that the covalent bonding strength of Pb-Br3 is stronger than that of Pb-Br4, as the SHG contribution of an individual anion can be weakened when it forms stronger covalent bonds with its surrounding cations[59].

Table 2. Contributions of the individual atoms to the largest SHG component $d_{25}$ of [R-MePEA][C3A]PbBr$_4$, and [R-MePEA][C4A]PbBr$_4$. $A_\tau$ is the contribution (in %) from a single atom $\tau$, and C$_A$ that from all atoms of the same type. $^{VB}A_\tau$ is the contribution (in %) of the VBs, and $^{CB}A_\tau$ from the CBs. The contributions from the *s*, *p*, and *d* states of the atom $\tau$ to of $^{VB}A_\tau$ and $^{CB}A_\tau$ are also shown. W$_A$ refers to the number of the same type of atoms (i. e. on the same Wyckoff site) in a unit cell. As the $A_\tau$ for C, N, and H atoms are very small, only the total averaged value is given.

| Material | Atom | W$_A$ | A$_\tau$ | C$_A$ | $^{VB}A_\tau$ | $^{CB}A_\tau$ | $^{VB}_sA_\tau$ | $^{VB}_pA_\tau$ | $^{VB}_dA_\tau$ | $^{CB}_sA_\tau$ | $^{CB}_pA_\tau$ | $^{CB}_dA_\tau$ |
|---|---|---|---|---|---|---|---|---|---|---|---|---|
| [R-MePEA][C3A]PbBr$_4$ | Pb | 2 | 19.9 | 39.8 | 7.2 | 12.7 | 5.3 | -1.0 | 3.0 | 0.1 | 12.1 | 0.5 |
| | Br1 | 2 | 5.3 | 10.7 | 1.4 | 3.9 | -1.6 | 3.0 | 0.0 | 1.0 | 1.8 | 1.1 |
| | Br2 | 2 | 4.3 | 8.7 | 0.6 | 3.7 | -1.6 | 2.2 | 0.0 | 1.1 | 1.6 | 1.1 |
| | Br3 | 2 | 1.8 | 3.5 | 2.4 | -0.6 | -0.2 | 2.6 | 0.0 | -0.2 | -0.9 | 0.5 |
| | Br4 | 2 | 5.8 | 11.6 | 5.9 | -0.2 | -0.1 | 6.1 | 0.0 | -0.3 | -0.4 | 0.5 |
| | C | 24 | 0.4 | 8.7 | 0.5 | -0.1 | 0.0 | 0.5 | 0.0 | 0.2 | -0.3 | 0.0 |
| | N | 4 | 1.4 | 5.6 | 0.5 | 0.9 | 0.0 | 0.5 | 0.0 | 0.1 | 0.8 | 0.0 |
| | H | 48 | 0.2 | 11.4 | 0.0 | 0.2 | 0.0 | 0.0 | 0.0 | 0.1 | 0.1 | 0.0 |
| [R-MePEA][C4A]PbBr$_4$ | Pb | 2 | 26.4 | 52.7 | 5.2 | 21.2 | 3.9 | -0.8 | 2.1 | 0.1 | 20.7 | 0.3 |
| | Br1 | 2 | 7.4 | 14.8 | 2.5 | 4.9 | -0.9 | 3.4 | 0.0 | 1.5 | 2.5 | 1.0 |
| | Br2 | 2 | 8.3 | 16.6 | 3.2 | 5.1 | -0.9 | 4.1 | 0.0 | 1.4 | 2.7 | 1.0 |
| | Br3 | 2 | -2.6 | -5.2 | -1.9 | -0.7 | -0.1 | -1.7 | 0.0 | -0.2 | -0.8 | 0.3 |
| | Br4 | 2 | 2.3 | 4.7 | 2.9 | -0.6 | 0.0 | 2.9 | 0.0 | -0.2 | -0.5 | 0.2 |
| | C | 26 | 0.2 | 6.1 | 0.5 | -0.3 | 0.2 | 0.3 | 0.0 | 0.1 | -0.4 | 0.0 |
| | N | 4 | -0.3 | -1.3 | -0.7 | 0.4 | 0.0 | -0.7 | 0.0 | 0.1 | 0.3 | 0.0 |
| | H | 52 | 0.2 | 11.6 | 0.1 | 0.1 | 0.1 | 0.0 | 0.0 | 0.1 | 0.0 | 0.0 |

According to the individual atomic contribution to the SHG response, the contribution of an atomic group can be calculated by summing the contributions of the center atom and those of its ligands. Therefore, the group contribution of PbBr$_4$ to SHG response is calculated as follows,

$$\chi^{(2)}_{[PbBr_6]} = \chi^{(2)}_{Pb} + 2 \times \left(\chi^{(2)}_{Br1}/2\right) + 2 \times \left(\chi^{(2)}_{Br2}/2\right) + 1 \times \chi^{(2)}_{Br3} + 1 \times \chi^{(2)}_{Br4}.$$
(1)

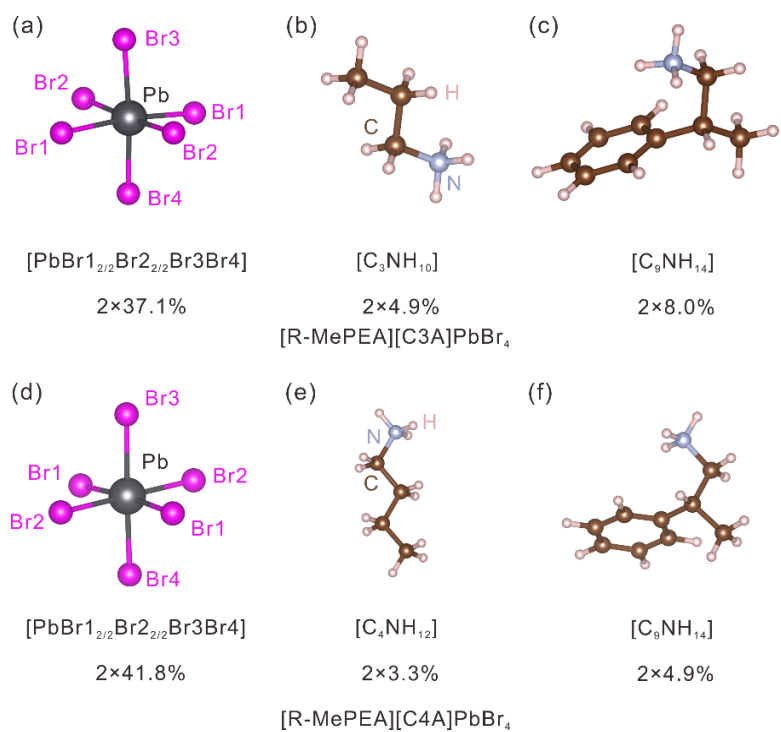

**Figure 3.** Group contributions to the largest second harmonic generation component $d_{25}$ of [R-MePEA][C3A]PbBr$_4$ (a-c), and [R-MePEA][C4A]PbBr$_4$ (d-f).

The total group contribution for the largest second harmonic generation component $d_{25}$ of [R-MePEA][C3A]PbBr$_4$, and [R-MePEA][C4A]PbBr$_4$ are shown in Figure 3, which shows that the metal-centered groups [PbBr$_6$] contribute much more strongly to the SHG response than the organic ligand. That is, the total contribution of the inorganic part for [R-MePEA][C3A]PbBr$_4$, and [R-MePEA][C4A]PbBr$_4$ are ~72.3 and 83.6%, which far surpasses that of the organic part (i.e., ~25.7, 16.4%), respectively. Our results confirm that the inorganic layer contributes dominantly to SHG response, while the organic part is important in the stabilization of the crystal structure. Thus, although the mixing of chiral and achiral organic cations, contributes to 'chiralize' the entire crystal, thus activating the SHG response, the further optimization of the NLO properties is mainly due to the electronic properties of the framework. At the same time, this suggests that they may additionally be modified by exploiting the tuning of the organic cations, which are affecting the properties of the framework through hydrogen-bonding-induced structural distortions, i.e. chirality transfer.

## 4. Conclusion

The structural relaxation, electronic structures, and nonlinear optical properties of alkyl-aryl mixed cation 2D HOIPs, [R-MePEA][C3A]PbBr$_4$, and [R-MePEA][C4A]PbBr$_4$, were studied using the first-principles calculations based on DFT. The electronic band gaps were calculated at different approximations, e.g. standard DFT, SOC, and HSE. We find that the presence of Br⋯NH noncovalent interactions is important in the formation of the organic and inorganic hybrid compounds. The calculated DOS, band structures, and COHP show that the top VBs are dominated by nonbonding Br 4p states, while the bottom CBs are made up of unoccupied Pb 6p states and minority C 2p orbitals. The static effective SHG value of [R-

MePEA][C3A]PbBr$_4$, and [R-MePEA][C4A]PbBr$_4$ were calculated as 0.493, and 0.173 pm/V respectively (i.e., ~1.49 and 0.53 ×$d_{eff}^{KDP}$) which are relatively strong compared with other lead-containing HOIPs. Besides, these two compounds also show moderate birefringence value (Δn) at 1064 nm. The relatively wide band gaps, SHG responses, and moderate birefringence of the two OIHMHs might suggest they are suitable for applications in nonlinear optical materials. Furthermore, the total contribution of the inorganic part for [R-MePEA][C3A]PbBr$_4$, and [R-MePEA][C4A]PbBr$_4$ are ~72.3 and 83.6%, which far surpasses that of the organic part (i.e., ~25.7, 16.4%), respectively, reflecting that the inorganic part contributes dominantly to SHG response, while the organic part is important in the stabilization of the crystal structure. Our study suggests that mixed organic chiral 2D HOIPs are potentially good candidates for NLO properties. Further study in this direction is required, both from the experimental and theoretical side in order to enhance or optimize their NLO responses.

## Supporting Information

This material is available free of charge via the Internet at http://pubs.acs.org.

## AUTHOR INFORMATION

**Corresponding Authors**

* Shuiquan Deng (sdeng@fjirsm.ac.cn)

* Alessandro Stroppa (alessandro.stroppa@spin.cnr.it)

**Author Contributions**

A.S. and S.D. proposed the project. X.C., S.M. and H.M. performed the calculations. The manuscript was written through the contributions of all authors.

‡ These authors contributed equally to this work.
## Notes

The authors declare no competing financial interest.

## ACKNOWLEDGMENT

The work at FJIRSM was financially supported by the National Natural Science Foundation (NSF) of China (22031009, 21921001, 22075282); the National Key R& D Program of China (2021YFB3601501); the NSF of Fujian Province (2023J01212). The work has been performed under Project HPC-EUROPA3 (HPC17Y3BP5 and HPC17P14RF), with the support of the EC Research Innovation Action under the H2020 Programme; CNR-SPIN, the Italian National Research Council Institute for Superconductors, with HPC resources and support provided by Cineca. We acknowledge support from the Italian Ministry of Research under the PRIN 2022 Grant No 2022F2K7J5 with title Two dimensional chiral hybrid organic-inorganic perovskites for chiroptoelectronics PE 3 funded by PNRR Mission 4 Istruzione e Ricerca - Component C2 - Investimento 1.1, Fondo per il Programma Nazionale di Ricerca e Progetti di Rilevante Interesse Nazionale PRIN 2022 - CUP B53D23004130006. A.S. thanks the support from *bilateral project CNR e NSFC (Cina), "Ferroelectric and chiral hybrid organic-inorganic perovskites", biennio 2024-2025").*

**Supporting Information**

**1. Partial response functional (PRF) method**[1-2]

The contribution of a certain occupied energy region between $E_B$ and valence band maximum (VBM), $\zeta_V(E_B)$, to each SHG coefficient $\chi^{(2)}_{ijk}$ is determined by considering only those excitations from all occupied states between $E_B$ and VBM to all the unoccupied states of the conduction bands (CBs), and the contribution, $\delta\zeta_V(E_B)$, of specific occupied states of energy $E_B$ to each $\chi^{(2)}_{ijk}$ by the excitations from that energy to all unoccupied states of the CBs.

$$\delta\zeta_V(E_B) = -\frac{d\zeta_V(E_B)}{dE_B} \qquad (1)$$

Similarly, the contribution, $\zeta_C(E_B)$, of a certain unoccupied region between conduction band minimum (CBM) and $E_B$ to each $\chi^{(2)}_{ijk}$ is determined by the excitations from all occupied states of the VBs only to all unoccupied states between CBM and $E_B$, and the contribution, $\delta\zeta_C(E_B)$, of specific unoccupied states of energy $E_B$ to each $\chi^{(2)}_{ijk}$ by the excitations from all occupied states of the VBs only to that energy.

$$\delta\zeta_C(E_B) = \frac{d\zeta_C(E_B)}{dE_B} \qquad (2)$$

**2. Atom response theory (ART) analysis**

To evaluate the individual atom contributions to the SHG components, $d_{ij}$, it is computationally more convenient to express the corresponding PRFs in terms of the band index $I_B$, $\zeta(I_B)$,[1] where the band index $I_B$ runs from 1 to $N_{tot}$ (i.e., the total number of band orbitals) with increasing energy, $E_B$, from $E_{min}$ to $E_{max}$. Here, $\zeta_V(I_B)$ and $\zeta_C(I_B)$ are denoted as $^{VB}\zeta_j$ and $^{CB}\zeta_j$, respectively, with $I_B$ replaced by a subscript $j$.

Suppose that a specific atom $\tau$ has $L$ atomic orbitals with a coefficient $^{VB}C^{\vec{k}j}_{L\tau}$ in the valence band $j$ at a wave vector $\vec{k}$. The total contribution $^{VB}A_\tau$ of an atom $\tau$ makes to the SHG coefficient from all the VB bands $j$ is written as

$$^{VB}A_\tau = \frac{\Omega}{(2\pi)^3}\int d\vec{k} \cdot \sum_{L,j} {}^{VB}\zeta_j \left|{}^{VB}C^{\vec{k}j}_{L\tau}\right|^2 \qquad (3)$$

where $\Omega$ is the unit cell volume, $^{VB}\zeta_j$ is the corresponding PRFs in terms of the band index j. Similarly, the total contribution $^{CB}A_\tau$ of an atom $\tau$ makes to the SHG coefficient from all the CB bands $j$ is written as

$$^{CB}A_\tau = \frac{\Omega}{(2\pi)^3}\int d\vec{k} \cdot \sum_{L,j} {}^{CB}\zeta_j \left|{}^{CB}C^{\vec{k}j}_{L\tau}\right|^2 \qquad (4)$$

in which we assumed that the atom has $L$ atomic orbitals with coefficient $^{CB}C^{\vec{k}j}_{L\tau}$ in the conduction band $j$ at a wave vector $\vec{k}$. To calculate the actual contribution of each constituent atom in a unit cell to the total SHG response, one needs to consider the signs of $^{VB}\zeta_j$ and $^{CB}\zeta_j$.

The total contribution, $A_\tau$, each individual atom makes to the SHG response from both the VBs and the CBs (i.e., from all the bands) is given by

$$A_\tau = \frac{\left({}^{VB}A_\tau + {}^{CB}A_\tau\right)}{2} \qquad (5)$$

where the factor of 1/2 is applied to remove the double counting of each excitation.

**Table S1.** The comparison of the experimental crystal structure with the calculated crystal structure, and the overall similarity, showing that the experimental and calculated structures are essentially coincident.

| Compound | Method | Degree of lattice distortion (s) | Maximum distance $d_{max}$ (Å) | Arithmetic mean value $d_{av}$ (Å) | Measurement of similarity ($\Delta$) |
|---|---|---|---|---|---|
| [R-MePEA][C3A]PbBr$_4$ | PBE-GGA | 0.00 | 0.6920 | 0.1842 | 0.028 |
| | DFT-D2 | 0.00 | 0.8260 | 0.2475 | 0.034 |
| [R-MePEA][C4A]PbBr$_4$ | PBE-GGA | 0.00 | 0.5020 | 0.1857 | 0.027 |
| | DFT-D2 | 0.00 | 0.6250 | 0.2476 | 0.034 |

**Table S2.** Bond length values after structure optimization for [R-MePEA][C3A]PbBr$_4$ and [R-MePEA][C4A]PbBr$_4$, as well as the experimental values.

| Bonds | [R-MePEA][C3A]PbBr$_4$ Bond lengths (Å) | | | [R-MePEA][C4A]PbBr$_4$ Bond lengths (Å) | | |
|---|---|---|---|---|---|---|
| | Experimental[3] | PBE-GGA | DFT-D2 | Experimental[3] | PBE-GGA | DFT-D2 |
| Pb1-Br1 | 2.993 | 3.018 | 3.034 | 3.010 | 3.040 | 3.076 |
| Pb1-Br2 | 3.000 | 3.021 | 3.061 | 3.025 | 3.045 | 3.072 |
| Pb1-Br3 | 3.037 | 3.069 | 3.081 | 3.008 | 3.045 | 3.071 |
| Pb1-Br4 | 3.009 | 3.049 | 3.086 | 2.995 | 3.034 | 3.053 |
| Pb1-Br5 | 2.925 | 2.982 | 3.075 | 2.924 | 2.970 | 3.063 |
| Pb1-Br6 | 3.093 | 3.113 | 3.213 | 3.089 | 3.127 | 3.227 |
| C-C | 1.390 | 1.394 | 1.398 | 1.363 | 1.396 | 1.399 |
| N-C | 1.480 | 1.495 | 1.520 | 1.481 | 1.498 | 1.523 |
| N-H | 0.890 | 1.048 | 1.040 | 0.889 | 1.036 | 1.032 |

**Table S3.** Specific optical rotation (in deg dm$^{-1}$g$^{-1}$cm$^3$) at B3LYP/6-311++G(d,p) for MePEA enantiomers.

| Molecule | λ (nm) | (R)-Neutral | (R)- Cationic |
|---|---|---|---|
| MePEA | 1064 | 23,9 | -12,7 |
| | 850 | 38,9 | -20,5 |
| | 700 | 60,4 | -31,5 |
| | 633 | 76,5 | -39,8 |
| | 589 | 91,2 | -47,3 |
| | 578 | 95,5 | -49,5 |
| | 546 | 110,3 | -57,1 |
| | 436 | 204,1 | -108,4 |
| | 365 | 361,8 | -223,8 |
| | 355 | 399,9 | -261,8 |
| | 300 | 814,1 | -2505,4 |
| | 250 | 2966,0 | 820,1 |
| | 200 | 861626,4 | -17875.9 |
| | 150 | 581,4 | -175210.5 |

**Table S4.** The electronic transition energies (in units of nm and eV), oscillator strengths (*f*), transition electric dipole moments ($M_{el.}$, in Debye), transition magnetic dipole moments ($M_{magn.}$, in Bohr magneton, BM), computed rotatory strength ($R_{vel.}$ and $R_{lengt.}$ are in 10$^{-40}$ erg-esu-cm) for MePEA enantiomers.

| Molecule | λ (nm) | ΔE (eV) | F | $M_{el.}$ | $M_{magn.}$ | $R_{vel.}$ | $R_{lengt/.}$ |
|---|---|---|---|---|---|---|---|
| | Neutral | | | | | | |
| (R)-MePEA | 235 | 5.2700 | 0.0026 | 0.362 | 0.500 | 1.4510 | 1.5512 |
| | 232 | 5.3478 | 0.0197 | 0.984 | 1.191 | 1.2690 | 1.6971 |
| | 218 | 5.6964 | 0.0012 | 0.238 | 0.324 | -1.9253 | -1.7214 |
| | 217 | 5.7096 | 0.0386 | 1.335 | 0.524 | 8.2412 | 8.2060 |
| | 214 | 5.8037 | 0.0114 | 0.719 | 0.757 | -2.2997 | -2.0164 |
| | 209 | 5.9385 | 0.0099 | 0.662 | 0.869 | -12.2122 | -12.6104 |
| | Cationic | | | | | | |
| (R)-MePEA$^+$ | 293 | 4.2326 | 0.0185 | 1.074 | 0.224 | -1.5595 | -1.6071 |
| | 289 | 4.2906 | 0.0020 | 0.351 | 0.136 | -0.2946 | -0.2781 |
| | 234 | 5.3090 | 0.0014 | 0.260 | 0.126 | -0.6615 | -0.6679 |
| | 219 | 5.6594 | 0.0152 | 0.842 | 0.514 | 1.6136 | 1.4565 |
| | 216 | 5.7434 | 0.0048 | 0.469 | 0.301 | -0.9600 | -0.9518 |
| | 214 | 5.8027 | 0.0323 | 1.211 | 0.087 | -1.1468 | -1.1670 |

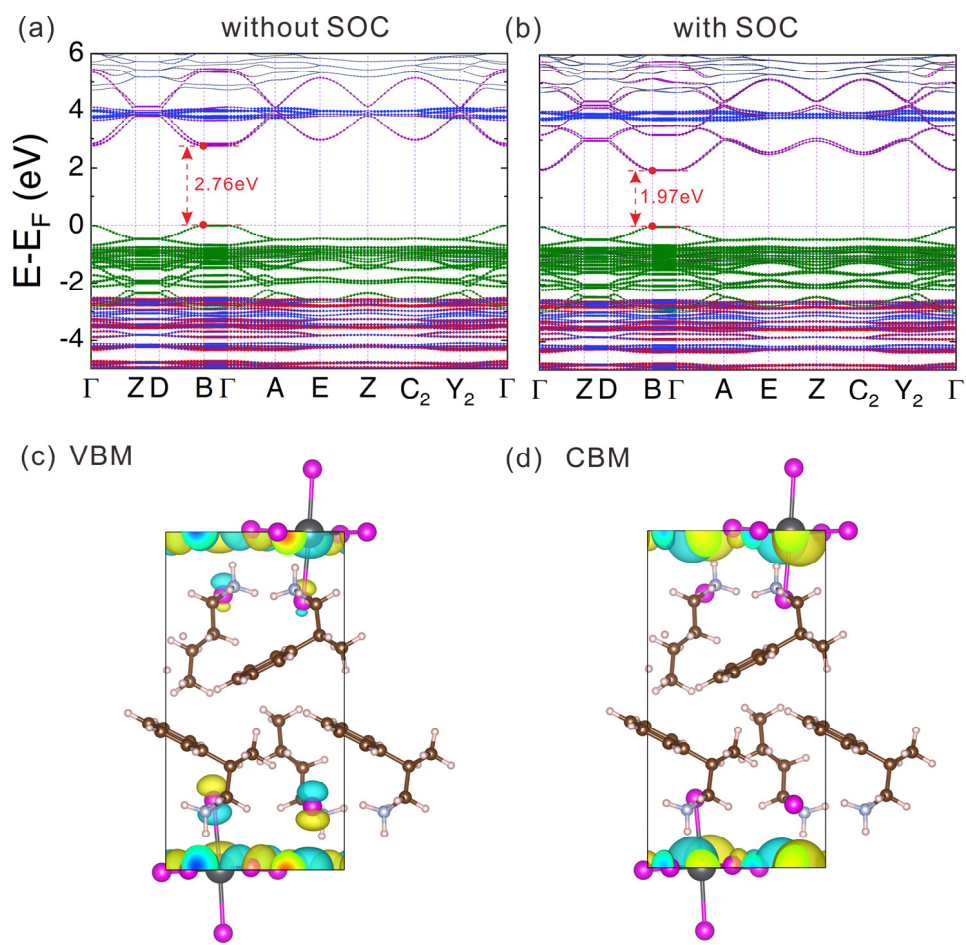

**Figure S1.** Band structures of [R-MePEA][C4A]PbBr$_4$ (a, b) without and with SOC. Γ=(0, 0, 0), Z=(0, 0.5, 0), D=(0, 0.5, 0.5), B=(0, 0, 0.5), A=(-0.5, 0, 0.5), E=(-0.5, 0.5, 0.5), C$_2$=(-0.5, 0.5, 0), Y$_2$=(-0.5, 0, 0). Key: green=Br 4p, purple=Pb 6p, blue=C 2p, red=H p. Wavefunction of VBM and CBM at B point for [R-MePEA][C4A]PbBr$_4$ (c, d).

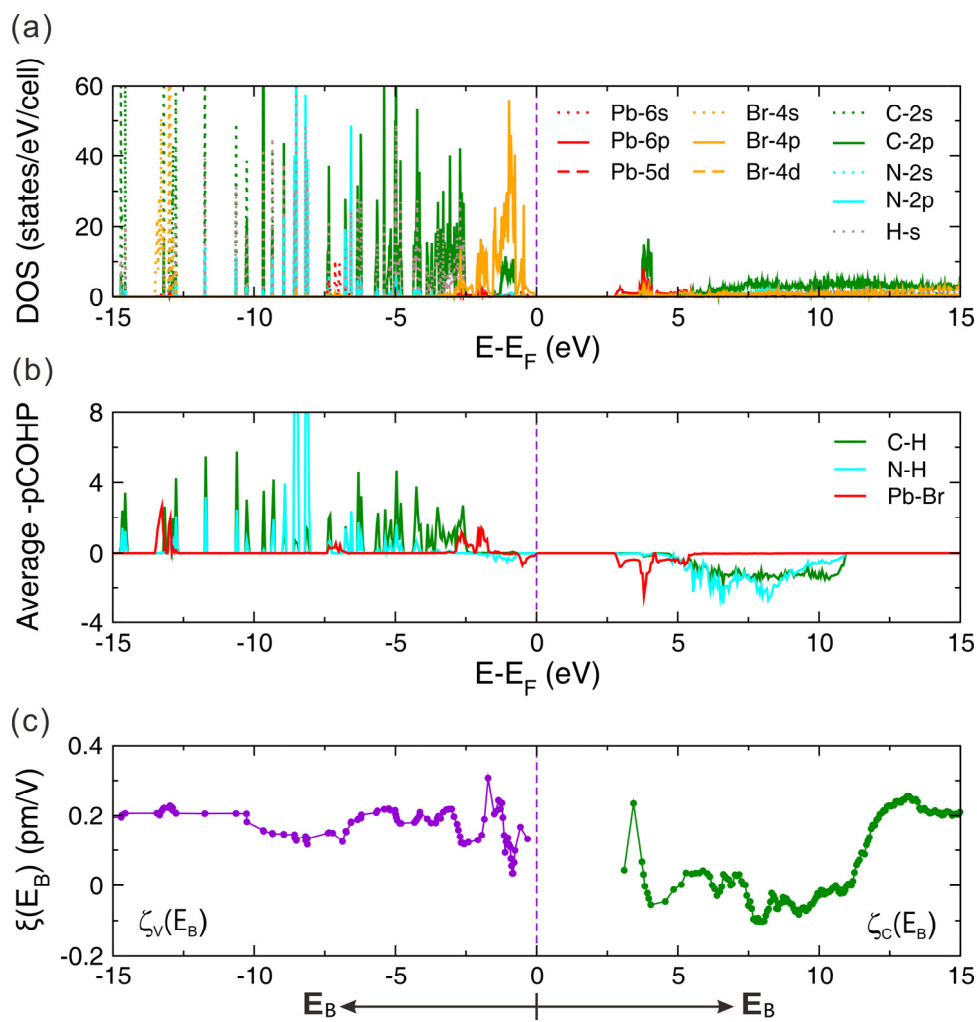

**Figure S2** Calculated PDOS (a), average -pCOHP (b), and $\zeta(E_B)$-vs-$E_B$ plot for the largest SHG component $d_{25}$ (c) of [R-MePEA][C4A]PbBr$_4$.

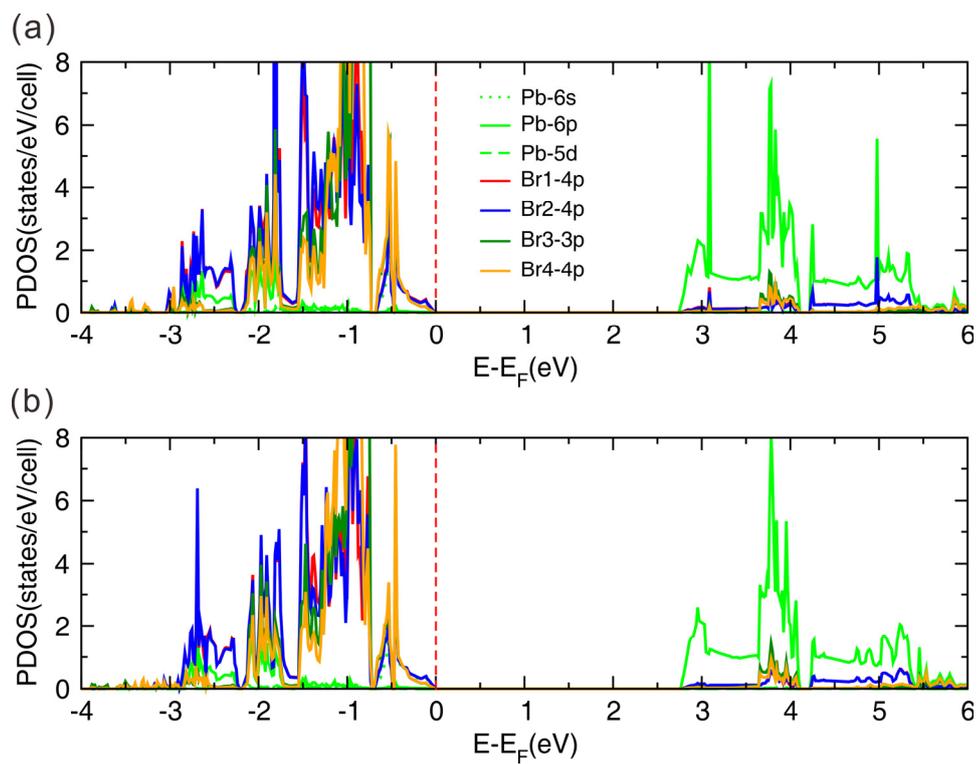

**Figure S3** PDOS of the inorganic part (Pb, Br1, Br2, Br3 and Br4 atoms) of [R-MePEA][C3A]PbBr$_4$ (a), and [R-MePEA][C4A]PbBr$_4$ (d).

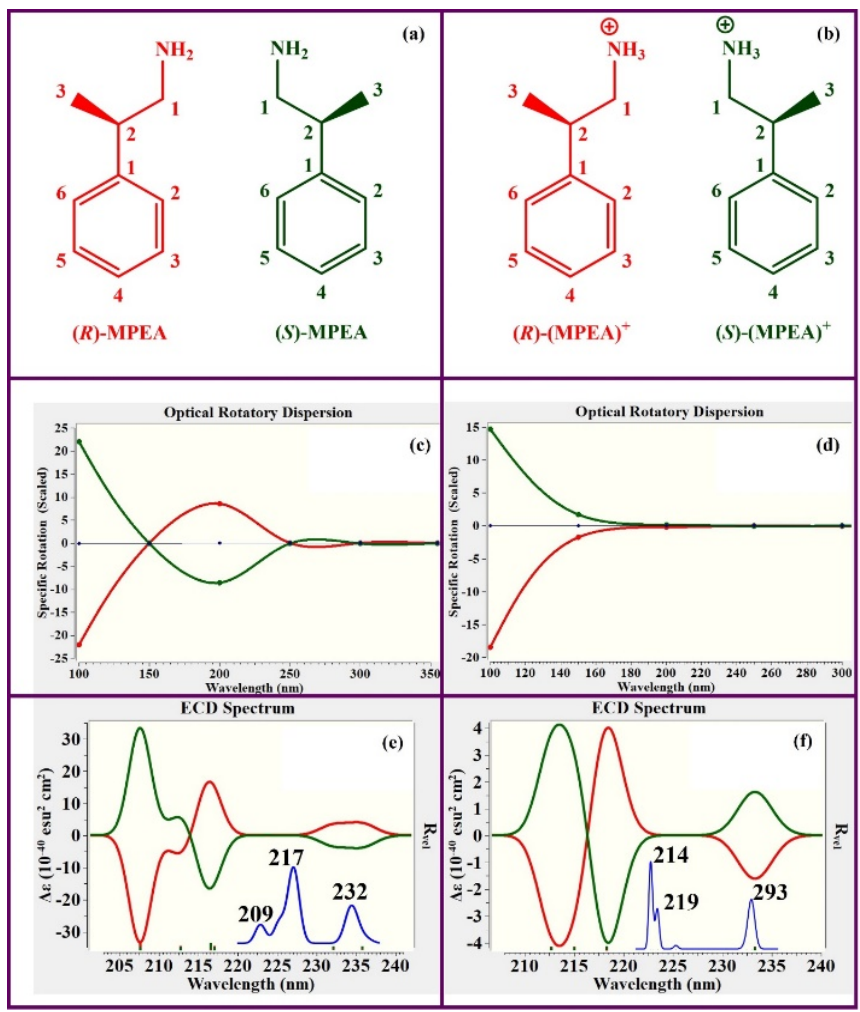

**Figure S4.** (a) Graphical representation of the neutral MePEA enantiomers; (b) cationic MePEA enantiomers; (c) ORD spectra (neutral); (d) ORD spectra (cation); (e) ECD spectra (neutral); (f) ECD spectra (cation). Specific optical rotation degrees (in deg dm-1g-1cm3 unit) are scaled by $10^{-5}$. The red curves are for (R)-MePEA and the green curves are for (S)-MePEA enantiomers.